\documentclass{article}

\usepackage{PRIMEarxiv}

\usepackage[linktocpage=true]{hyperref}       
\usepackage{booktabs}       
\usepackage{amsfonts}       
\usepackage{nicefrac}       
\usepackage{microtype}      
\usepackage{fancyhdr}       
\usepackage{graphicx}       
\graphicspath{{media/}}     

\usepackage{amsmath,amsthm,amssymb,scrextend}
\usepackage{etoolbox}
\usepackage{appendix}
\AtBeginEnvironment{appendices}{}
\usepackage{enumerate}
\usepackage{csquotes}
\usepackage{subcaption}
\usepackage{newtxmath}
\usepackage{changepage}
\usepackage{bm}

\usepackage{rotating}

\usepackage{bbm}

\usepackage{enumerate}

\usepackage{chapterbib}

\hypersetup{colorlinks=true,linkcolor=blue,anchorcolor=blue,citecolor=blue,filecolor=blue,urlcolor=blue,bookmarksnumbered=true,pdfview=FitB}

\usepackage{xcolor}

\title{Forecasting Influenza Hospitalizations Using a Bayesian Hierarchical Nonlinear Model with Discrepancy
}

\author{
  Spencer Wadsworth \\
  University of Connecticut \\
  Storrs, Connecticut\\
  \texttt{iac25002@uconn.edu} \\
   \And
  Jarad Niemi \\
  Iowa State University \\
  Ames, Iowa\\
  \texttt{niemi@iastate.edu} \\
}

\begin{document}
\maketitle
\begin{abstract}
The annual influenza outbreak leads to significant public health and economic 
burdens making it desirable to have prompt and accurate probabilistic forecasts 
of the disease spread. The United States Centers for Disease Control and 
Prevention hosts annually a national flu forecasting competition which has 
led to the development of a variety of flu forecast modeling methods. For the 
first several years of the competition, the target to be forecast was weekly 
percentage of patients with an influenza-like illness (ILI), but in 2021 the 
target was changed to weekly hospitalization counts. Reliable state and national 
hospitalization data has only been available since 2021, but for ILI the data 
has been available since 2010 and has been successfully forecast for several 
seasons.
In this manuscript, we introduce a two component 
Bayesian modeling framework for 
forecasting weekly hospitalizations utilizing both hospitalization data and ILI 
data. The first component is for modeling ILI data using a nonlinear 
Bayesian hierarchical model. This component includes modeling discrepancy 
between a nonlinear functional model and the observed ILI which requires 
thoughtful decisions
for modeling constraints and selecting prior parameter distributions.
The second component is for modeling 
hospitalizations as a function of ILI. For hospitalization forecasts, ILI is 
first forecasted and then hospitalizations forecasts are produced using
ILI forecasts 
as a linear or quadratic predictor. The combination of the models is done in a
manner similar to using the Bayesian cut.
In a simulation study, two ILI forecast 
models, including one similar to the winning model for two seasons of the CDC 
forecast competition from Osthus et al.
%
and a nonlinear Bayesian hierarchical model from Ulloa 
are compared with other standard forecasts models.
The usefulness of including a systematic model 
discrepancy term in the ILI model is specifically addressed and proves 
to greatly improve forecasts. In a real data analysis, 
forecasts of state and national 
hospitalizations for the 2023-24 flu season are made and compared with forecasts
from 20 other competing models. The forecasts from the proposed methodology 
outperformed all but one of 20 competing models. 
\end{abstract}
\section{Introduction}

Every year the seasonal influenza outbreak burdens the public health system by 
infecting millions, causing an influx of primary care visits and 
hospitalizations and leading to between 290,000 and 650,000 deaths worldwide 
\cite[]{whoflufact2023}. Molinari et al. \cite[]{molinari2007annual} estimated 
the United States' annual economic burden from medical costs, loss of income, 
and deaths to be over \$87 billion. Accurate forecasting of infectious diseases 
can inform public decision making and ease the burden of an outbreak 
\cite[]{turtle2021accurate, lutz2019applying}.
There is a growing consensus that disease forecasts should be probabilistic in 
nature \cite[]{gneiting2014probabilistic, bracher2021evaluating}, and it has 
been shown that reporting forecast uncertainty along with predictions may lead 
to better decision making 
\cite[]{ramos2013probabilistic, joslyn2012uncertainty, winkler1971probabilistic}.


To better inform public decision making regarding the flu epidemic, in 2013 the 
United States Centers for Disease Control and Prevention (CDC) organized a 
national flu forecasting competition, also known as FluSight 
\cite[]{biggerstaff2016results, mathis2024evaluation,cdcfluforecasting2024}. 
The competition lasted from the fall of 2013 through spring of 2014, and
originally over a dozen teams of researchers from academic and industry 
backgrounds participated in FluSight by contributing their own forecast models. 
Besides the 2020-21 season, FluSight has been operated annually and 
researchers outside 
the CDC have been invited to participate. From hereon, we denote a 
flu season by the year in which the outbreak began, i.e. the 2020 season 
season outbreak began in 2020 and lasted until spring 2021.
Initially the target data for 
forecasts was influenza-like illness (ILI) data. ILI is the proportion of 
patients who meet a healthcare provider and who display flu like symptoms, 
and ILI data has been available at the state and national level since the 2010 
flu season \cite[]{cdcfluview2023,cdc2024fluviewportal}. 
The collaborative ILI forecasting effort has led to a number of modeling 
developments in flu forecasting 
\cite[see references therein for more examples]{mcandrew2021adaptively, 
osthus2021multiscale, osthus2019dynamic, ulloa2019}, and in their 
introduction, Osthus et al.  \cite[]{osthus2019dynamic} categorized the most 
commonly used flu forecasting models into four classes including mechanistic 
models based on differential equation compartmental models, agent based models 
based on population simulation, machine learning/regression models including 
data driven machine learning and statistical models, and data assimilation 
models which are constructed by assimilating mechanistic models into a 
probabilistic framework. An additional forecast modeling class used in FluSight 
involves the combination of several forecasts into a single ensemble forecast, 
which has been shown to perform well relative to individual models 
\cite[]{mcandrew2021adaptively, ray2020ensemble, yamana2016superensemble}.

The administration of FluSight saw few changes during the first seven seasons
between 2013 and 2019, 
but the onset of the COVID-19 pandemic and subsequent developments for COVID-19 
forecasting led to major modifications. As a result of the COVID-19 pandemic 
which began during the 2019 flu season, the typical flu outbreak behavior was 
altered between the 2019 and 2022 seasons \cite[]{mathis2024evaluation}. 
The COVID-19 pandemic led to the creation of the Health and Human Services 
(HHS) Patient Impact and Hospital Capacity Data System 
\cite[]{healthdata2024covidts} which contains COVID-19 and flu hospitalization 
data, and the COVID-19 Forecast Hub was founded. The COVID-19 Forecast Hub 
was based on FluSight but with certain major adjustments including how the 
forecast uncertainty is represented and the addition of the weekly publication 
of a multi-model ensemble forecast as the official forecast of the CDC 
\cite[]{bracher2021evaluating, Cramer2022-hub-dataset}. Using estimated 
quantiles for representing forecast uncertainty and creating a multi-model 
ensemble are both aspects of the COVID-19 Forecast Hub which were adopted by 
FluSight. Additionally the target of the flu forecasts 
changed from being ILI data to being HHS hospitalization data, which reports 
the number of hospitalizations due to a laboratory confirmed flu infection  
\cite[]{mathis2024evaluation,healthdata2024covidts}. This in part was 
as a result of 
having COVID-19 cases in the population making ILI data, already only a proxy 
for flu behavior, more difficult to interpret.

The contribution of this manuscript is to introduce a two component 
Bayesian modeling framework 
for modeling HHS hospitalization forecasts where hospitalization data and 
years of ILI data are used to inform forecast models. The first modeling 
component is a hierarchical model of ILI data. 
The second is a model of hospitalization 
data with ILI as a predictive covariate.
Herein we use Bayesian ILI models similar to those in Osthus et al. 
\cite[]{osthus2019dynamic} and Ulloa \cite[]{ulloa2019} for ILI forecasting.
The model of Osthus et al. \cite[]{osthus2019dynamic} is a combined data 
assimilation and statistical regression model which involves a compartmental 
model in a probabilistic framework. Their model also includes an additional 
component for capturing a systematic discrepancy between the deterministic 
part of the model and the actual data, an idea which was first introduced by 
Kennedy and O'Hagan \cite[]{kennedy2001bayesian}. The model in Ulloa 
\cite[]{ulloa2019} is a Bayesian hierarchical regression model with an 
underlying function intended to capture the trajectory of the seasonal 
ILI data. Herein, we provide a framework under which discrepancy modeling may 
be used along with a general function modeling ILI data. Modeling discrepancy
introduced challenges which required thoughtful selection of model 
constraints and prior distributions to be used effectively.

In line with the newer FluSight standard of forecasting hospitalizations, we 
model hospitalizations as a linear function of ILI. 
The flu hospitalization forecasts are a linear mapping of ILI forecasts
allowing for ILI data from many seasons to be exploited
and assist in forecasting hospitalizations.
To produce hospitalization forecasts, a scheme similar to 
the Bayesian cut is used where forecasts from the posterior predictive
distribution of an ILI model are plugged into the hospitalization model
to produce a posterior predictive hospitalization forecast distribution.
We show that this scheme produces accurate and reasonably well calibrated
probabilistic forecasts.
The forecasts produced in this manuscript are compared with out of the box 
forecast models in a simulation study, and in a real data analysis the 
forecasts are compared with 20 other forecasts from the 2023 FluSight forecast
competition. The proposed models outperform both the out of the box methods, and
they outperform all but one of the FluSight models.

In section \ref{sec:data} we review the ILI and hospitalization data provided 
by the CDC and targeted by FluSight. 
In section \ref{sec:functions} the modeling framework contributed by this 
manuscript is given. In the same section, functions similar to those used by 
Osthus et al. \cite[]{osthus2019dynamic} and Ulloa \cite[]{ulloa2019} are 
defined.
 These functions are the susceptible-infectious-recovered (SIR) compartmental 
 model and the asymmetric Gaussian (ASG) function  respectively. Model fitting 
 and implementation are described at the end of the section. Section 
 \ref{sec:simulation2} includes a simulation study where four ILI forecast 
 models and their use in forecasting hospitalizations are compared. Commonly 
 used proper scoring rules \cite[]{gneiting2007strictly} are used for 
 comparing the forecasts. Forecasting of the 2023 flu outbreak along with 
 assessment and comparison of 20 FluSight models is performed in 
 section \ref{sec:analysis}. 
 Finally, the manuscript is concluded in section \ref{sec:conclusion} with 
 general observations and some discussion.

 \section{Flu outbreak data} \label{sec:data}
In this section we introduce, define, and visually evaluate
ILI and hospitalization 
data. ILI and hospitalization data have been the object 
of forecasting for FluSight with ILI being the target for the first seven
seasons, 2013 to 2019, and hospitalizations being the target since the 
2022 season. Both of 
these data were collected at the state, territorial, and national levels and 
were reported at least weekly. Overall the data is reported for 53 locations 
including the 50 US states, the District of Columbia (DC), Puerto Rico (PR), 
and the US national level. We will refer to forecast targets throughout this 
manuscript. A target is the specific horizon, 1, 2, 3, or 4-weeks ahead, for a 
specific location and week during the season.  

\subsection{Influenza-like illness data}
The US Outpatient Influenza-like Illness Surveillance Network collects 
information on respiratory illness from outpatient visits to health care 
providers. Each week, over 3,400 outpatient health care providers in all 50 
US states, PR, 
DC, and the US Virgin Islands report the total number of outpatient 
visits along with the number of ILI cases. An ILI case is defined as a 
``fever (temperature of $100^{\circ}$F[$37.8^{\circ}$C] or greater) and a cough 
and/or a sore throat.`` Prior to the 2021 season, the definition included 
``without a known cause other than influenza`` \cite[]{cdcfluview2023}. Because 
other illnesses such as COVID-19, RSV, and the common cold may induce similar 
respiratory symptoms, ILI may include patients infected with some disease 
other than influenza. To know whether or not a sick patient is infected with 
influenza requires a laboratory test. 

In 2013, when FluSight began, the ILI data was the object of the forecasts. 
The data was released publicly at HHS region levels, and forecast teams were 
asked to provide forecasts of several ILI targets on the regional levels 
including season onset, 1-4 week ahead ILI levels, and the week of peak ILI 
activity \cite[]{biggerstaff2016results,mcgowan2019collaborative}. Currently, 
the ILI data is collected by the CDC and published on an online portal for 
viewing at the national, HHS region, census, and state levels 
\cite[]{cdc2024fluviewportal}. To obtain ILI data, we used the \texttt{R} 
package \texttt{cdcfluview} which provides functions for downloading the data 
\cite[]{rudis2021cdcfluview}. Weekly ILI data from the national, HHS region, 
and census levels are available from the 1997 flu season up to the current 
season. At the state level, data is available from the 2010 flu season to the 
current season. 

\begin{figure}[hbt!]
    \centering
    \includegraphics[scale=.5]{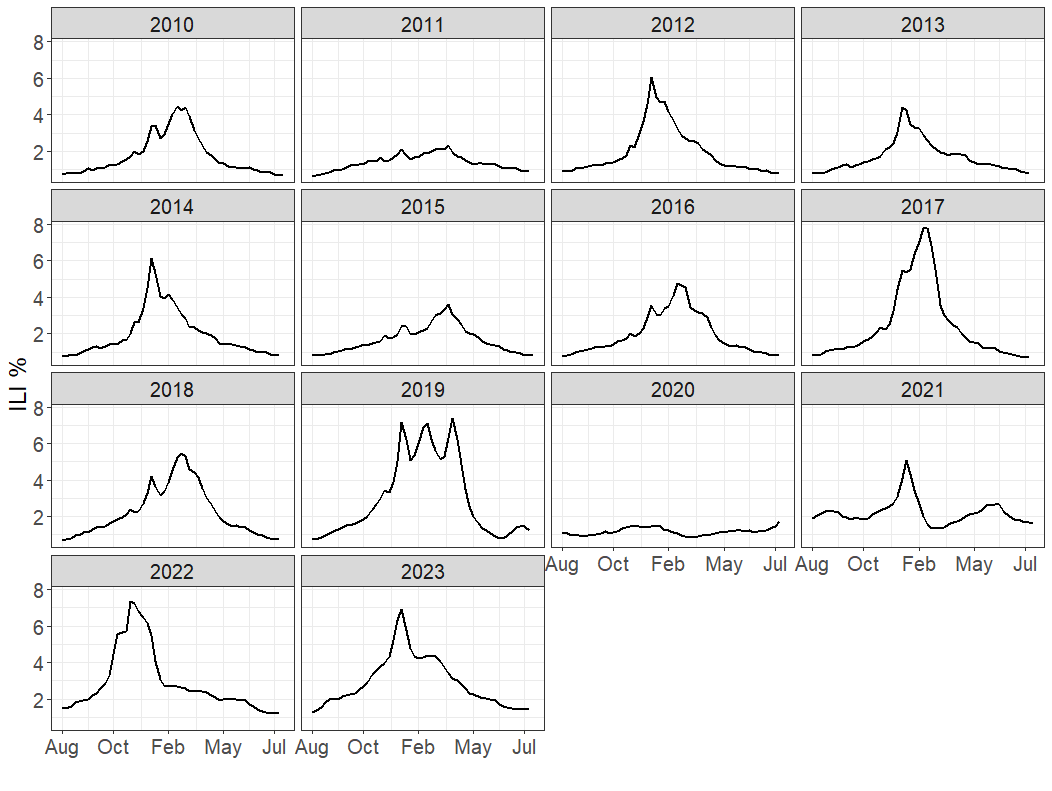}
    
    \centering
    \includegraphics[scale=.42]{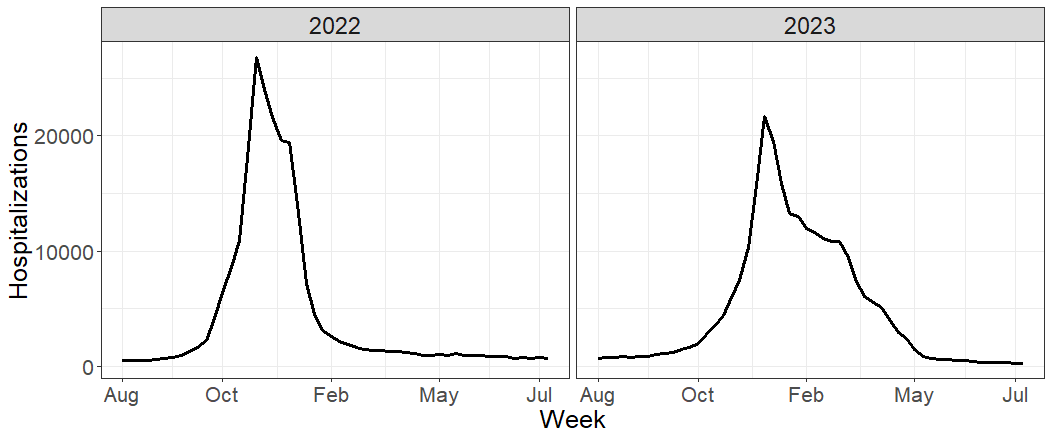}
    \caption{Percentage of outpatient visits with an influenza-like illness 
    (ILI) in the US for seasons 2010 to 2023 (top) and weekly flu confirmed 
    hospitalization counts at the national level for 2022 and 2023 flu seasons 
    (bottom). Week 1 is the first week of August of the year the flu season 
    begins and the last week of the season is the last week of July of the 
    following year.}
    \label{fig:us_ili}
\end{figure}

The top of figure \ref{fig:us_ili} shows the ILI data at the national level for 
flu seasons 2010 to 2023. For most seasons there are 52 weeks listed, but for 
the 2010, 2015, and 2021 seasons there are 53 weeks because those
seasons had 53 Sundays. 
To better align with the flu behavior, week 1 is set as the first week of 
August and week 52 or 53 is the last week in July of the following year. For 
example, week 1 of the 2013 season corresponds to the first week of August 2013, 
and week 52 of the same season corresponds to the last week of July 2014. This 
convention is used for the remainder of this manuscript. 

Notable from the plots in the top of figure \ref{fig:us_ili} is the regular 
trajectory of the ILI. With the exception of season 2020, the ILI begins low 
at week 1 and increases as the fall and winter progress until the ILI reaches 
a peak. As spring progresses to summer, the ILI decreases to low values. As 
Osthus et al. \cite[]{osthus2019dynamic} point out, there is nearly always 
either a global or local peak at week 22 which typically corresponds to the 
week between Christmas and New Year's day. 
Whether local or global, the ILI holiday peak is generally expected and thought 
to be due to widespread holiday travel, school closure, or other unique social 
behavior   
\cite[]{ewing2017contact, garza2013effect}.
The only seasons when there was not a peak at week 22 were 2022, when 
the season peak occurred particularly early, and 2020 which was greatly 
influenced by the COVID-19 pandemic.

\subsection{Hospitalization Data}

Weekly hospital admission data, which was used as the object of FluSight 
forecasting for the 
2022 and 2023 seasons, is based on the CDC's National Healthcare Safety 
Network (NHSN) dataset entitled \textit{HealthData.gov COVID-19 Reported 
Patient Impact and Hospital Capacity by State Timeseries}. 
In February 2022 it became mandatory for 
all hospitals to report the number of COVID-19 and influenza hospitalizations, 
and since then reporting of hospitalizations has become widespread. These data 
were updated every Wednesday and Friday according to NHSN guidelines 
\cite[]{healthdata2024covidts}. Weekly influenza hospitalization data is
defined as the number of newly hospitalized patients with a confirmed diagnosis 
of influenza. From hereon, influenza hospitalizations
will be referrred to as hospitalizations.

The bottom of figure \ref{fig:us_ili} shows the weekly national 
hospitalizations for the 2022 and 2023 flu seasons. These plots show a 
trend similar to the ILI plots in that during the early 
weeks of the season hospitalizations are low, but they increase in the fall 
to a peak and then decrease until the flu outbreak ends. For both the 
2022 and 2023 seasons, the hospitalizations peaked during the same week as 
ILI, and in 2023 that peak occurred during the holiday week 22. Additional 
plots of ILI data and hospitalization
data for all 50 US states are included in the supplementary material 
\cite[]{wadsworth2024bas}. These plots show flu outbreak trajectories 
similar to those in figure \ref{fig:us_ili}.


\section{ILI and hospitalization forecast modeling} \label{sec:functions}


The typical behavior of the ILI data to begin low, rise to a peak, and then 
fall motivates the use of a nonlinear function which follows 
a similar trajectory. Compartmental models are standard mathematical models 
used for modelling disease outbreaks which may capture ILI behavior. 
One important compartmental model is the 
susceptible-infectious-recovered (SIR) 
disease transmission model, which has been used by some to model 
ILI data \cite[]{osthus2019dynamic, allen2017primer}.  
The SIR is a mechanistic model used to capture transmission of a particular
disease, and since ILI may be influenced by more 
diseases than influenza, the disease transmission described by the SIR model
can only partially contribute to the true ILI data-generating mechanism
\cite[]{osthus2019dynamic}.

Ulloa \cite[]{ulloa2019} 
chose to use the asymmetric Gaussian (ASG) function to model ILI data.
The ASG is a smooth function which may approximate the 
same shape as the ILI trajectory.
The ASG function has more parameters and is thus slightly more flexible than
the SIR model, and thus may be able to roughly capture ILI.
However, there may still be 
systematic behavior which it cannot capture. Thus when using either the SIR
or ASG models, an additional component for modeling discrepancy should be 
included.

In the first part of this section, we present an ILI model 
similar to the model in Osthus et al. \cite[]{osthus2019dynamic}. With some 
generalization, the model may incorporate any appropriate linear or
nonlinear function, 
though the focus here is on the SIR and ASG functions, defined later
in this section. 
With the aim of forecasting hospitalizations, we also introduce a linear 
model of hospitalization data with ILI as a predictive covariate. To forecast 
hospitalizations, ILI data is first forecasted and the forecast is then,
as a covariate,
plugged into the hospitalization model to produce
hospitalization forecasts.

Because we desire probabilistic forecasts, Bayesian modeling is a 
natural framework for describing forecast uncertainty. 
Though our main interest is in effeciently producing a probabilistic 
forecast and not 
necessarily in estimating model parameters, certain modeling decisions and 
prior distribution selections proved critical in producing good 
forecasts. After defining the ILI and hospitalization models, 
we describe the 
selection of prior distributions, model fitting, and producing forecasts
from the posterior predictive distribution by combining ILI and hospitalization
models using a scheme similar to the 
Bayesian cut \cite[]{plummer2015cuts}.

\subsection{ILI Model} \label{sec:ili_model}

The proposed model for ILI for any location is given in (\ref{eq:ili_model}). 
Here $ILI_{s,w}$ is the ILI for flu season $s$ and week $w = 1, 2, ..., W$, 
where $W = 52$ or $W = 53$, depending on how many Sundays there are in a given 
season. The ILI is a proportion, so the Beta random variable is a natural 
selection for modeling. Under the parameterization of the Beta distribution 
used in (\ref{eq:ili_model}) the expected value is $\pi_{s,w}$ and the 
variance is $\pi_{s,w}(1 - \pi_{s,w})/(1 + \kappa_s)$, making $\kappa_s$ a 
scale parameter. The nonlinear function $f_{\theta_s}(w)$ captures the 
trajectory of the ILI, and $\gamma_w$ is a discrepancy term shared by 
all seasons which captures 
systematic patterns not captured by $f_{\theta_s}(w)$. The additional 
discrepancy term $\upsilon_{s,w}$ captures additional season 
specific patterns.

\begin{equation}
\begin{aligned}
    \label{eq:ili_model}
        ILI_{s,w} &\overset{ind}{\sim} \text{Beta}(\pi_{s,w}\kappa_s,\; \kappa_s(1 - \pi_{s,w})) \\
        \text{logit}(\pi_{s,w}) &= f_{\theta_s}(w) + \gamma_w + \upsilon_{s,w}
\end{aligned}
\end{equation}
In Osthus et al. \cite[]{osthus2019dynamic} 
$f_{\theta_s}(w) = \text{logit}(I_{s,w})$ where $I_{s,w}$ is the infectious 
compartment of the SIR model from (\ref{eq:sir_diff}) in section 
\ref{sec:sir_func}. In Ulloa \cite[]{ulloa2019}, 
$f_{\theta_s}(w) = ASG_{\theta}(w)$ from (\ref{eq:asg_function}) in section 
\ref{sec:asg_func}. In Ulloa \cite[]{ulloa2019}, modeling is done hierarchically 
over seasons. Herein we also adopt the hierarchical modeling so as to borrow 
information contained in the data across all seasons.

Model (\ref{eq:ili_model}) was fit 
independently for each location. This decision was made largely due to the 
limited time given by FluSight to produce forecasts. That is,
during the competition, data were typically released on Wednesday afternoons and
forecasts were due by the evening of the same day giving only a few hours to 
produce forecasts. A natural extension of the model would be to allow for states
to borrow information from neighboring states.

\subsection{Susceptible-Infectious-Recovered (SIR) compartmental model} \label{sec:sir_func}

\begin{figure}[hbt!]
    \centering
    \includegraphics[scale=.6]{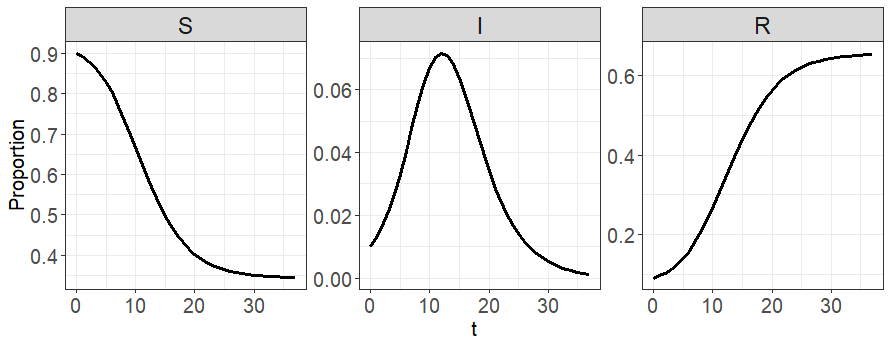}
    \caption{Susceptible-infectious-recovered (SIR) model separated by 
    compartments. The three compartments are the susceptible compartment 
    (left), infectious (center), and recovered (right). In this 
    example, $S_0/\rho > 1$.}
    \label{fig:sir_traj}
\end{figure}

The SIR compartmental model is a mathematical model used for modeling disease 
outbreaks and was introduced in 1927 by Kermack and McKendrick 
\cite[]{kermack1927contribution}. Since then, compartmental models have 
become standard in modeling infectious diseases 
\cite[]{allen2008mathematical}, and many extensions have been made and 
studied \cite[for example]{simon2020sir, allen2017primer, van2008deterministic}. 
The SIR mathematical model includes three compartments and assumes that at 
any time $t>0$ every individual in a closed population belongs to exactly 
one compartment. The three compartments are susceptible ($S$), infectious ($I$), 
and recovered ($R$), and their interaction over the course of an outbreak is 
described by the differential equations in (\ref{eq:sir_diff}).    

\begin{equation}
    \label{eq:sir_diff}
    \frac{dS}{dt} = -\beta SI, \quad \frac{dI}{dt} = \beta S I - \delta I, \quad \frac{dR}{dt} = \delta I
\end{equation}
Here $S$, $I$, and $R$ represent the proportion of the population in each 
compartment such that $S + I + R = 1$ for all $t$. The trajectory is 
determined by the disease transmission rate $\beta > 0$ and the recovery 
rate $\delta > 0$. Respectively, these may be thought of as the expected 
proportion of susceptible individuals who will be infected by an infectious 
individual and the expected rate of recovery to an immune state for a newly 
infected individual. Whether or not a disease outbreak is classified as an 
epidemic is determined by the initial susceptible population $S_0$, or the 
susceptible population at time $0$, and the parameter $\rho = \delta/
\beta$. If $S_0/\rho > 1$, the outbreak is considered an epidemic. It is 
non-epidemic if $S_0/\rho \leq 1$ \cite[]{osthus2019dynamic}. Figure 
\ref{fig:sir_traj} shows the trajectory of the three compartments of an SIR 
model where $S_0$ and $\rho$ were selected to match an outbreak that is an 
epidemic. In the case where $S_0 \leq \rho$, the trajectory for the $I$ 
compartment will be to never increase. The increase to a peak and subsequent 
decrease in the $I$ compartment of figure \ref{fig:sir_traj} illustrate why
it is 
reasonable to use the SIR model to partially capture ILI, although it is
unlikely to completely capture the true data generating mechanism. Thus the 
use of modeling discrepancy.

\subsection{Asymmetric Gaussian (ASG) function} \label{sec:asg_func}
\begin{figure}[hbt!]
    \centering
    \includegraphics[scale=.45]{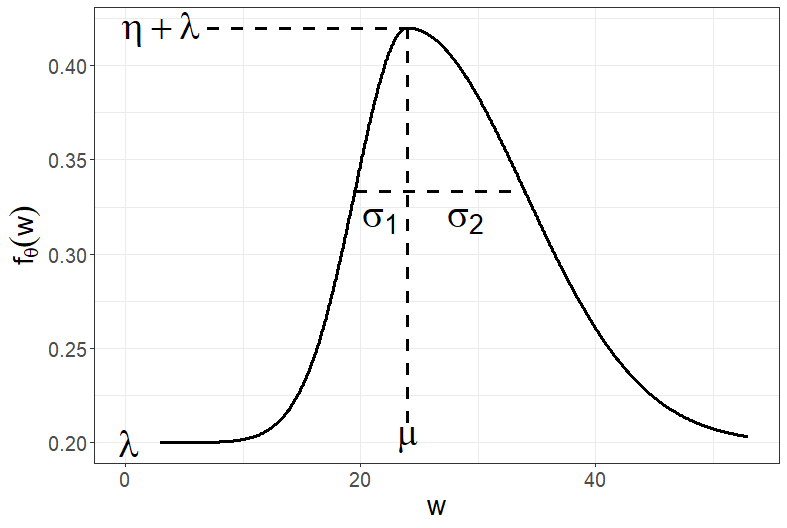}
    \caption{Example plot of asymmetric Gaussian (ASG) function showing the 
    shape of the function in relation to the parameters 
    $\lambda, \; \eta, \; \mu, \; \sigma_1, \; \text{and} \; \sigma_2$}
    \label{fig:asg_function}
\end{figure}

The ASG function is another example of a nonlinear function which can 
approximate the trajectory of the flu outbreak. The ASG was previously used 
by Ulloa to model and forecast ILI \cite[]{ulloa2019}. It has also been used 
to model vegetation growth and satellite sensor data 
\cite[]{lewis2020extracting, jonsson2002seasonality, hird2009noise, 
beck2006improved, atkinson2012inter}. The ASG is a modification of the 
asymmetric Gaussian distribution \cite[]{wallis2014two} and is characterized 
by its rise to a peak and fall from that peak, which rise and fall 
may not occur at the same 
rate, as shown in figure \ref{fig:asg_function}. The ASG function is denoted 
as $ASG_\theta(w)$ where 
$\theta = (\lambda, \nu, \mu, \sigma_1^2, \sigma_2^2)$, $\nu > 0$, 
$\lambda > 0$, $\mu \in (-\infty, \infty)$, $\sigma_1, \sigma_2 > 0$ and 
$w \in (1, ..., W)$ is week. The function is defined in 
(\ref{eq:asg_function}).

\begin{equation}
    \label{eq:asg_function}
    ASG_{\theta}(w) = 
    \begin{cases}
        \lambda + (\nu - \lambda) \text{exp}[-(w - \mu)^2/2\sigma^2_1], \;\;\; w < \mu \\
        \lambda + (\nu - \lambda) \text{exp}[-(w - \mu)^2/2\sigma^2_2], \;\;\; w \geq \mu
    \end{cases}
\end{equation}
In this manuscript, we use a slightly reparameterized version of function
(\ref{eq:asg_function}) shown in (\ref{eq:asg_function_rep}), 
where $\eta = \nu - \lambda > 0$. 
This constraint guarantees that the function has a peak greater than $\lambda$.

\begin{equation}
    \label{eq:asg_function_rep}
    ASG_{\theta}(w) = 
    \begin{cases}
        \lambda + \eta \text{exp}[-(w - \mu)^2/2\sigma^2_1], \;\;\; w < \mu \\
        \lambda + \eta \text{exp}[-(w - \mu)^2/2\sigma^2_2], \;\;\; w \geq \mu
    \end{cases}
\end{equation}

\subsection{Model discrepancy}

\begin{figure}[hbt!]
    \centering
    \includegraphics[scale=.5]{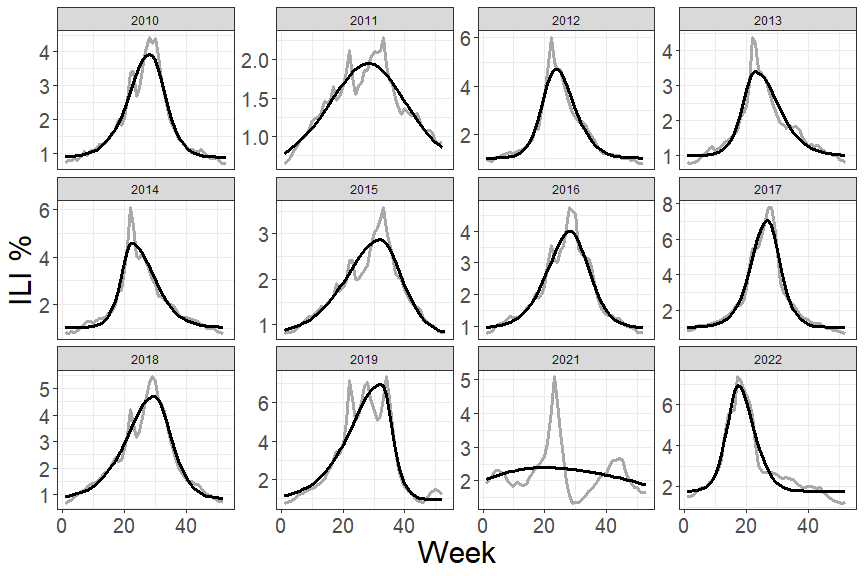}
    \caption{Observed US national influenza-like illness (ILI) percentage for 
    seasons 2010 to 2022 excluding 2020 (grey) overlaid with MLE of an 
    asymmetric Gaussian (ASG) model for the ILI data (black)}
    \label{fig:asg_fits}
\end{figure}

\begin{figure}[hbt!]
    \centering
    \includegraphics[scale=.45]{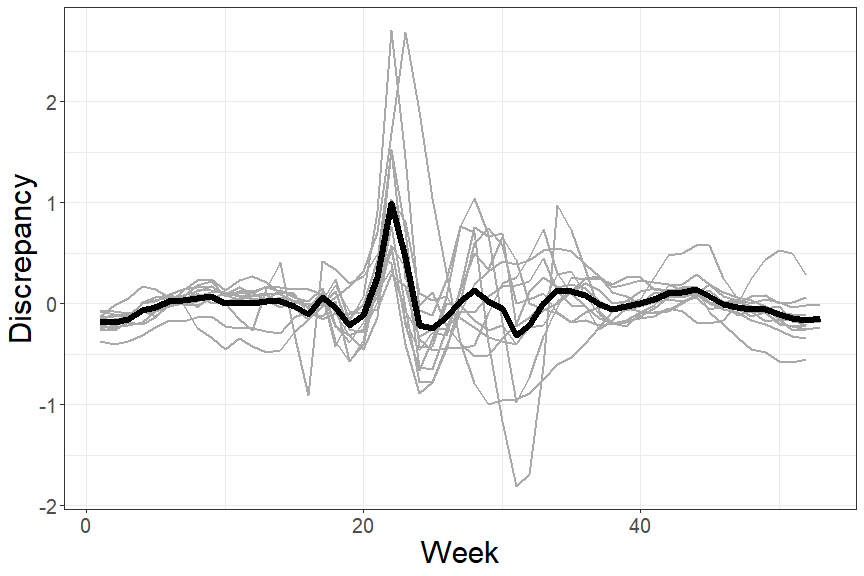}
    \caption{Difference between observed US national influenza-like illness 
    (ILI) and MLE fits for an asymmetric Gaussian (ASG) model for each season 
    2010 to 2022 excluding 2020 (grey) and the average difference of all 
    seasons (black)}
    \label{fig:discrepancy}
\end{figure}

The SIR and ASG functions are useful for capturing the main trend of the ILI 
data, but as Osthsu et al. \cite[]{osthus2019dynamic} point out there may be 
systematic behavior in the ILI that these or other possible functions may not 
capture. 
Figures \ref{fig:asg_fits} and \ref{fig:discrepancy} are used together to 
illustrate the systematic discrepancy from a fitted function to ILI  data.
Figure 
\ref{fig:asg_fits} shows the US ILI percentage for all flu seasons from 2010 
to 2022 excluding 2020 with a best fit ASG function plotted over the ILI. The 
fits for each season were made by computing the maximum likelihood estimate 
(MLE) of a model given the ILI data where we assume the ASG function is the 
mean parameter of a Beta distributed random variable. Figure 
\ref{fig:discrepancy} shows the discrepancy between the model fit and the data 
for the same seasons. The grey lines show the difference between the data and 
the functions from figure \ref{fig:asg_fits} for each season, and the black 
line is the average discrepancy over all seasons. The lines show that the ASG 
function typically underpredicts week 22 and overpredicts week 23. Perhaps 
for other weeks, week 30 for example, there tends to be additional systematic 
behavior not captured by the ASG function.

Modeling discrepancy has been used in uncertainty analysis of 
simulators to capture systematic differences between mathematical models and 
reality \cite[]{ma2022multifidelity,brynjarsdottir2014learning,
arendt2012improving,kennedy2001bayesian}. Modeling discrepancy can lead to 
overfitting, particularly in forecasting scenarios, and may also lead to 
identifiability issues. 
Thus, care must be taken in setting parameter constraints as well as in the 
selection of prior distributions \cite[]{osthus2019dynamic,
brynjarsdottir2014learning}. 
For instance, when setting $f_{\theta_s}(w)$ to
be the ASG function
in model (\ref{eq:ili_model}), we found the parameter $\lambda$ from 
(\ref{eq:asg_function_rep}) difficult to 
estimate precisely because of lack of identifiability. This was resolved by 
assigning to $\lambda$ a more informative prior distribution.
Modeling of the discrepancy for ILI was done by Osthus et al. during the 2015 
and 2016 flu seasons where their model outperformed all others in the CDC flu 
forecasting challenge \cite[]{osthus2019dynamic}.
The terms $\gamma_w$ and $\upsilon_{s,w}$, where $s$ is the season and $w$ is the 
season week, are included in 
(\ref{eq:ili_model}) to capture the per week discrepancy between ILI and the 
function. Here $\gamma_w$ is shared across all seasons, whereas $\upsilon_{s,w}$
is specific to season $s$. To assist in estimating $\gamma_w$,
$\upsilon_{s,w}$ is only included for the current season being forecast and 
none of the previous seasons.
 As in Osthus's model, 
$\gamma_w$ and $\upsilon_{s,w}$ are modeled as reverse random walks, as shown 
in (\ref{eq:rw}). 
\begin{equation}
    \begin{aligned}
    \label{eq:rw}
        \gamma_w|\gamma_{w + 1} &\overset{ind}{\sim} N(\gamma_{w+1},\sigma^2_{\gamma}), \quad
        \gamma_{W} \sim N(0,\sigma^2_{\gamma_W}) \\
        \upsilon_{s,w}|\upsilon_{s, w + 1} &\overset{ind}{\sim} N(\upsilon_{s, w+1},\sigma^2_{\upsilon})
    \end{aligned}
\end{equation}

The idea for using the reverse random walk is that there are several previous 
seasons of ILI data, and assuming the random walk captures systematic behavior, 
fitting it hierarchically over seasons can assist in predicting future behavior 
in the current season. 
Reverse random walks have also been used with success in election forecasting 
and other flu forecasting models \cite[]{osthus2021multiscale, 
osthus2019dynamic, linzer2013dynamic}. The sum to zero constraint 
$-\gamma_1 = \sum_{w=2}^W \gamma_w$ is imposed on (\ref{eq:rw}) to improve 
identifiability as well as the constraint $\upsilon_{s, W} = 0$.

\subsection{Hospitalization model} \label{sec:hospital_model}

The second component for forecast modeling is based on the relationship between 
hospitalizations and ILI and is defined in (\ref{eq:final_hosp}).
This is an example of an autoregressive model with exogenous variables where 
the autoregressive lag is one (ARX(1)) 
\cite[]{raftery2010online,ljung1987system}. 

\begin{equation}
    \begin{aligned}
    \label{eq:final_hosp}
    H_{s,w} &= \alpha_{0s} + \alpha_{1s} (ILI_{s,w} \times P) + \alpha_{2s} 
    (ILI_{s,w} \times P)^2 + \phi H_{s,w-1} + \epsilon_{s,w}\\ 
    \epsilon_{s,w} &\overset{iid}{\sim} D_s(0, \sigma_{\epsilon_s}^2 \times P, 
    \omega_s) 
    \end{aligned}
\end{equation}

\begin{figure}[hbt!]
    \centering
    \includegraphics[scale=.2]{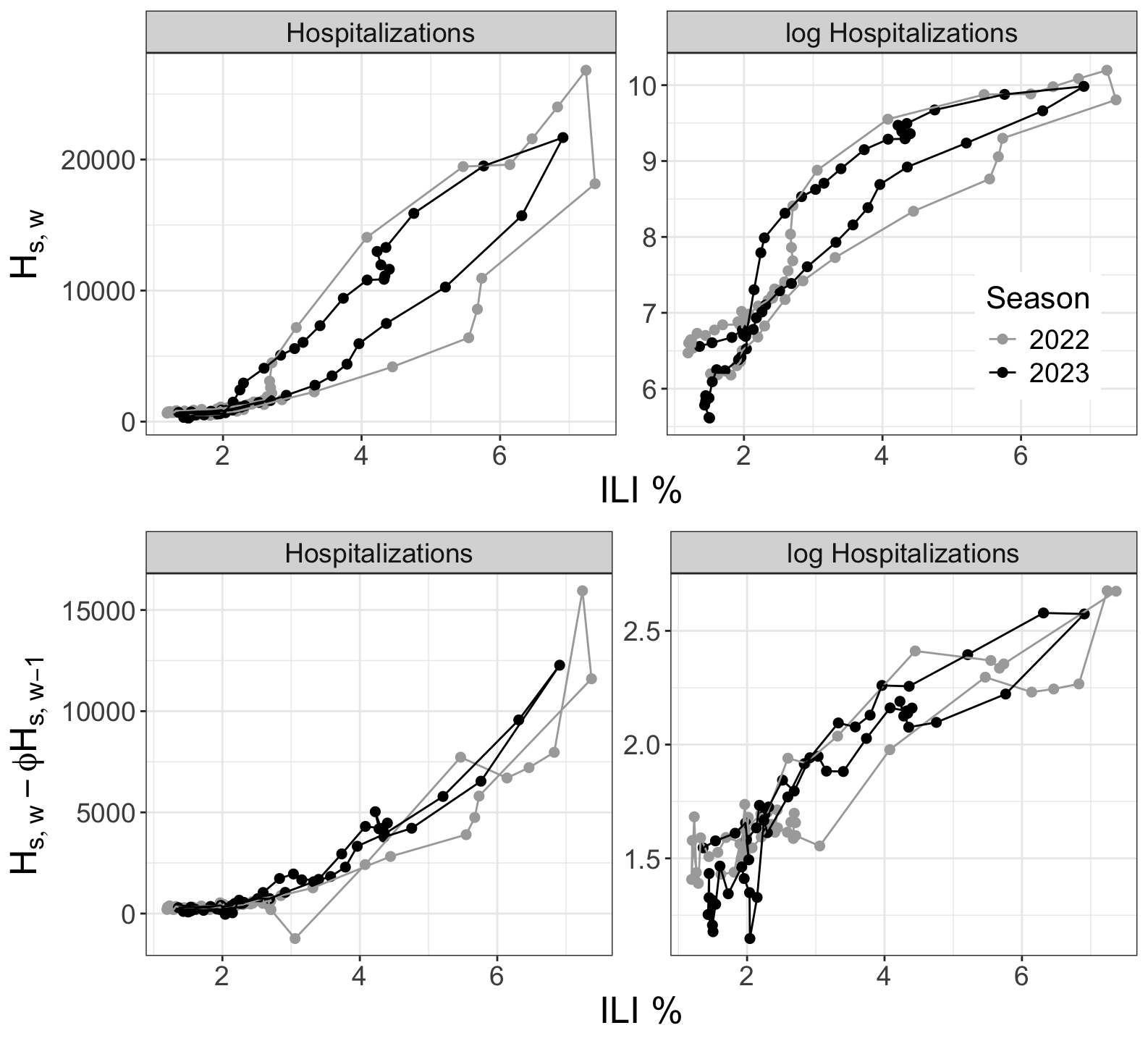}
    \caption{Scatterplots of national hospitalizations (left) or log 
    hospitalizations (right) lag differences by ILI \% where the lag is 
    scaled by $\phi$. Points are colored by season. For hospitalizations 
    $\phi = 0.6$, and for log hospitalizations $\phi = 0.77$. Lines through
    the points follow the weekly progression of influenza.}
    \label{fig:lag_scatter}
\end{figure}

Figure \ref{fig:lag_scatter} shows US national level data which helps
illustrate why this modeling decision was made.
The figure shows scatterplots of ILI percentage and 
hospitalizations on the top row.
The bottom row shows scatterplots of the difference between hospitalizations and 
1 week lags
by ILI percentage. The plots in the right column
are of the logarithm of hospitalizations. The lines through the points are
drawn by connecting hospitalizations for consecutive weeks during the flu 
season.
These plots show that a linear or quadratic relationship between ILI percentage 
and hospitalizations does not capture much of the temporal correlation in 
hospitalizations as shown by the obvious loops drawn in the plot. 
However, when including a 
single autoregressive lag of hospitalizations in the model, much of the time 
correlation is
accounted for as shown by the tighter and more frequent crossing of lines
in the bottom row. We found that models with more than 1 autoregressive
lag showed little or no forecast
improvement over 1 lag. 

In model (\ref{eq:final_hosp}), $H_{s,w}$ is the number of hospitalizations for 
week $w$ in season $s$, $\epsilon_{s,w}$ is an error term distributed 
according to some distribution $D_s$ with mean parameter 0, scale parameter 
$\sigma_{\epsilon_s}$, and the additional parameter $\omega_s$ which 
may be used as degrees of freedom
when $D_s$ belongs to the location-scale t (LST) family.

Like the ILI model (\ref{eq:ili_model}), 
the hospitalization model was fit independently for each location.
For fitting purposes $ILI_{s,w}$ is always 
multiplied by $P$ which is proportional to the population of the state or 
territory, in this case the total population divided by 50,000. This is done 
as a means of scaling so that the prior distribution assigned to  
$\boldsymbol{\alpha}_s = (\alpha_{0s}, \alpha_{1s}, \alpha_{2s})$, 
$\sigma_{\epsilon_s}$, and $\omega_s$ might reasonably be the same for all 
locations.

Like the ILI model, the hospitalization model in (\ref{eq:final_hosp}) is also 
fit via Bayesian posterior updating. To obtain forecasts for $H_{s, w^* + i}$, 
the ILI posterior predictive distribution is used along with the posterior 
distribution for the parameters in (\ref{eq:final_hosp}). 
We let $D_s$ belong to the normal family, though in the supplementary materials
we present a real forecast
analysis where we considered other distribution families
including the LST and lognormal families.


\subsection{Prior selection}

The prior distributions
selected for the ILI data model under both the SIR and ASG models 
largely follow the prior distribution 
selections in Osthus et al. \cite[]{osthus2019dynamic} 
and Ulloa \cite[]{ulloa2019} with a few exceptions where selecting
different priors improved 
identifiability, 
numerical stability, and/or we felt the adjusted prior made more sense for the 
problem. For model (\ref{eq:ili_model}), parameters which are common even when 
using different functions of $f_{\theta_s}(w)$ are $\kappa_s$, 
$\sigma_{\gamma}^2$, $\sigma_{\gamma_W}^2$, and $\sigma_{\upsilon}$. For the SIR function 
$\theta_s = (S_{0s}, I_{0s}, R_{0s}, \alpha_s, \rho_s)$, and for the ASG 
function $\theta_s = (\lambda_s, \eta_s, \mu_s, \sigma_{1s}^2, \sigma_{2s}^2)$. 
For the hospitalization model in (\ref{eq:final_hosp}) the parameter to be 
estimated is $\Psi = (\alpha_{0s}, \alpha_{1s}, \alpha_{2s}, \phi, \sigma_{\epsilon_s}, \omega_s)$.

The priors assigned were mostly noninformative, though in certain cases the 
prior distributions were selected for numerical stability as was the case 
for $\sigma_{\gamma}^2$ and $\sigma_{\gamma_W}^2$. For these two scale 
parameters only, rather than assigning a half-normal prior to the standard 
deviation parameters, as recommended by Gelman \cite[]{gelman2006prior}, the 
priors were assigned to the variance parameters. Univariate parameters were 
assigned either a normal distribution prior if the support was on $\mathbb{R}$,
a half-normal prior if the support is nonnegative, or a truncated-normal 
prior to match a more specific support. Under the ASG model, $\theta_s$ was 
modeled hierarchically over seasons so that for each season the transformed 
parameter $T(\theta_s) \sim N(\theta, \Sigma$) and priors distributions are 
assigned to $\theta$ and $\Sigma$ where
$T(\theta_s) = (\lambda_s, \text{log}(\eta_s), \mu_s, \text{log}(\sigma_{1s}^2), \text{log}(\sigma_{2s}^2))$.

Additional prior constraints were made to improve parameter identifiability. 
Just as was done in Osthus et al. \cite[]{osthus2019dynamic},
we set the initial value of the 
susceptible population compartment of the SIR model was set to $S_0 = 0.9$. 
The parameters $I_{0s}$, $\beta_s$, and $\rho_s$ were assigned informative 
priors. In a previous version of this manuscript, when fitting an ASG model
with discrepancy components,
rather than estimating all parameters through posterior updating,
identifiability was encouraged by setting
the parameter $\lambda_s$ from (\ref{eq:asg_function_rep})
to be equal to its MLE. In this updated manuscript, 
however, we use empirical Bayes and center the first component of the 
hierarichal prior $\theta$ to be the mean of MLEs for $\lambda_s$ across
all seasons and set the variance parameter so the prior was tight around the
mean. This led to improved mixing of posterior draws and better forecasts.
The 
specific prior distributions assigned to each parameter are listed in the 
supplementary materials.


\subsection{Hospitalization Forecasts Using the Bayesian Cut}

Rather than combining the ILI model in equation (\ref{eq:ili_model}) with the 
hospitalization model in equation (\ref{eq:final_hosp}) and fitting a single
posterior distribution, both models were fit separately and 1 to 4 week ahead
ILI forecasts from model (\ref{eq:ili_model}) were combined with the posterior
predictive distribution of model (\ref{eq:final_hosp}) to obtain 1 to 4 week
ahead hospitalization forecasts. In this scheme, the posterior distribution of 
model (\ref{eq:ili_model}) parameters was not directly influenced by the 
hospitalization data and the posterior distribution of model 
(\ref{eq:final_hosp}) parameters was not directly influenced by model 
(\ref{eq:ili_model}) parameters. Hence there was limited feedback between the
two components, but propagation of uncertainty in ILI forecasts was still 
permitted in the estimating of hospitalization forecasts.

This two-component modeling is remeniscent of the Bayesian cut 
described in \cite{plummer2015cuts} and \cite{nott2023bayesian}. Cutting a model
into two component models or ``modules`` may be justfied for several reasons
including to reduce time to fit the model, avoiding mixing issues of posterior
sampling distributions, improving predictive performance or to prevent
model misspecification in one model to influence the other 
\cite[]{nott2023bayesian, jacob2020unbiased, jacob2017better, plummer2015cuts}.

The decision to cut the forecast model into two component models
was made for several reasons.
When fitting the fully Bayesian joint model, the time to fit a single model using
Markov chain Monte Carlo (MCMC) sampling and
obtain forecasts was sometimes many hours whereas fitting the two models
separately took a fraction of the time. When assessing convergence of model
parameters, it was rare that posterior sampling chains for the joint model
showed convergence even after 55,000 iterations, whereas the component model 
fits gave no signs which raised concern for convergence. And most importantly,
the forecasts from the component models generally outperformed those of the 
joint model in terms of minimizing proper scoring rules.

Model (\ref{eq:ili_model}) was fit via Bayesian posterior updating. Future ILI 
forecasts were obtained via the posterior predictive distribution where for
week $w$, the predictive distribution was obtained by integrating over the 
parameters $\pi_s$, $\kappa_s$, $\sigma^2_{\gamma}$, and $\sigma^2_{\gamma_W}$ 
as in (\ref{eq:ili_post}) where 
$p(\boldsymbol{\pi}_s, \kappa_s, \sigma^2_{\gamma}, \sigma^2_{\gamma_W} | \textbf{ILI})$ 
is the density function of the posterior distribution for the model parameters. 
If the current week was $w$ then the desired forecasts were for weeks $w + i$ 
where $i$ is a positive integer. Here $\widetilde{ILI}_{s,w + i}$ represents the 
predicted or forecasted ILI at $i$ weeks in the future.

\begin{equation}
    \label{eq:ili_post}
    p(\widetilde{ILI}_{s,w +i} | \textbf{ILI}) = \int \int \int \int 
    p(\widetilde{ILI}_{s,w + i} | \boldsymbol{\pi}_s, \kappa_s, 
    \sigma^2_{\gamma}, \sigma^2_{\gamma_W}) p(\boldsymbol{\pi}_s, \kappa_s, 
    \sigma^2_{\gamma}, \sigma^2_{\gamma_W} | \textbf{ILI}) 
    d\boldsymbol{\pi}_s d \kappa_s d \sigma^2_{\gamma} d \sigma^2_{\gamma_W}
\end{equation}
Hospitalization forecasts $\widetilde{H}_{s,w + i}$ at week $w + i$ were then
obtained by combining the posterior predictive distribution of model 
(\ref{eq:final_hosp}) with (\ref{eq:ili_post}) and integrating over 
$\widetilde{ILI}_{s,w + i}$ as in (\ref{eq:hosp_post_forc}). Here,
$\boldsymbol{\alpha}$ represents all regression and variance parameters in
model (\ref{eq:final_hosp}).

\begin{equation}
\begin{aligned}
\label{eq:hosp_post_forc}
    p(\tilde{H}_{s,w + i} | \textbf{ILI}, \textbf{H}, \boldsymbol{\alpha}) &= 
       \int p(\tilde{H}_{s,w + i} | \textbf{H}, \textbf{ILI}, 
       \widetilde{ILI}_{s,w + i}, \boldsymbol{\alpha}) 
       p(\widetilde{ILI}_{s,w + i} | \textbf{ILI}) d\widetilde{ILI}_{s,w + i}
\end{aligned}
\end{equation}

\subsection{Parameter estimation and posterior predictive sampling}
\label{sec:implementation_posterior}

The models were fit via MCMC sampling using the 
\texttt{cmdstanr} package which was developed and is maintained by the Stan 
Development Team \cite[]{stan2024manual} \cite[]{gabry2022stan}. Stan 
implements Hamiltonian Monte Carlo (HMC) sampling with the No-U-turn 
sampler \cite[]{hoffman2014no}. The \texttt{cmdstanr} package provides 
several diagnostic statistics for assessing the sampler.
Plots of prior and posterior distributions for select parameters from ILI and 
hospitalization models are included in the supplementary material 
\cite[]{wadsworth2024bas}.

Model fit was assessed for four ILI models and for 
the hospitalization model. The ILI models included the SIR and 
ASG models and models with and without discrepancy modeling. When discrepancy 
is included, the models are denoted as SIRD and ASGD. These models were fit 
using US national data from 2010 to 2023 flu seasons, where data from the 
2020 season was excluded because of the unique behavior during that season. 
Sampling was done with four chains where from each chain 
60,000 posterior draws were sampled, and the first 10,000 draws were discarded 
as a burn-in. The $\hat{R}$ statistic \cite[]{vehtari2021rank} and the 
effective sample size ($ESS$) \cite[]{gelman2013bayesian} were calculated for 
each parameter, and for all models fit, the largest $\hat{R}$ was below the 
recommend threshold of 1.05, and the smallest $ESS$ was over 500.



To obtain forecast distributions of hospitalizations, draws from the posterior 
predictive distribution from the ILI model were used in conjunction with the 
posterior distribution of the hospitalizations model. When fitting model 
(\ref{eq:ili_post}), MCMC samples of $\widetilde{ILI}_{s,w:(w + 4)}$ were 
saved. Model (\ref{eq:final_hosp}) was fit and MCMC samples for the marginal 
distributions for the model parameters were saved. To obtain forecast 
distributions for $H_{s,w + i}$ where $i \in \{1,2,3,4\}$, the following 
steps were repeated $K$ times where $K$ is an integer for the number of desired 
samples. We set $K = 50,000$.

\begin{quote}
\begin{enumerate}[Step 1:]
  \item Sample $\widetilde{ILI}^*_{s,w:(w + 4)}$
  \item Sample $\alpha_{0s}^*$, $\alpha_{1s}^*$, $\alpha_{2s}^*$, $\phi^*$, 
  $\sigma^*_{\epsilon_s}$, $\omega^*_s$ from respective marginal posterior 
  distributions
  \item Sample $H^*_{s,w + i}$ from $D(\omega_s^*, \mu_{s, w + i}^*,\sigma^2_{\epsilon_s}$), where
  \begin{itemize}
    \item[] $\mu_{s,w + i}^* = \alpha_{0s}^* + \alpha_{1s}^* (ILI_{s,w + i}^* \times P) + \alpha_{2s}^* (ILI_{s,w + i}^* \times P)^2 + \phi^* H^*_{s,w + i - 1}$
  \end{itemize}
   \item Repeat step 3 for $i \in \{1,2,3,4\}$ to obtain $H^*_{s,(w + 1):(w + 4)}$
  \item Repeat steps 1-4 $K$ times
\end{enumerate}
\end{quote}
The sample $\{H^*_{s,w + i}\}^K$ was then used as the probabilistic forecast 
for hospitalizations at week $w + i$. 
Occasionally $\{H^*_{s,w + i}\}^K$ included a small number of negative values,
which do not make sense when the distribution is meant to forecast
hospitalizations, a nonnegative number. We attempted to alleviate this problem
by modeling log-hospitalizations or by assuming hospitalizations followed a
distribution truncated at 0, but these models led to poorer forecasts
and issues drawing from the posterior distribution.
For the forecast competition analysis in 
section \ref{sec:analysis}, all negative values of $\{H^*_{s,w + i}\}^K$ were 
set to 0 to reflect realistic values of hospitalizations and comply with the 
FluSight forecasting rules.

\section{Simulation Study} \label{sec:simulation2}
In this section, we present a simulation study conducted for comparing ILI 
models and further assessing the hospitalization forecast model. US ILI data 
is used, and hospitalization data is simulated. 
A leave-one-season-out (LOSO) approach was combined with a Monte Carlo 
simulation approach. For each replication, we simulated log-hospitalizations 
for all weeks during seasons 2010, ..., 2022, excluding 2020, using the 
existing ILI data as a predictive covariate. Each season was in turn 
``left-out`` and treated as if it was the most recent season for which forecasts
are desired. Fitting and forecasting was then done for weeks 14, 20, 26, 32, 
and 38 of the left out season, giving two weeks that tend to occur as flu 
cases increase, two as cases decrease, and one that occurs when the cases may 
be increasing or decreasing. Week 20 is a week leading up to the holiday week 
22 where ILI typically has a local peak. We were particularly interested in 
how important modeling discrepancy is for forecasting at week 20.

Besides the 4 ILI models, we also included forecasts for a naive baseline
(BASE) model and an autoregressive integrated moving average (ARIMA) model. 
The BASE model is a simple random walk with drift model, and the ARIMA model
is fit by selecting the best possible model give the data according to a 
fitting criterion. Both the BASE and ARIMA ILI models were fit using the 
\texttt{forecast} package in \texttt{R} \cite{hyndman2008forecast}.

For the simulation of hospitalizations, the parameters 
$\boldsymbol{\alpha}_s = (\alpha_{0s}, \alpha_{1s}, \alpha_{2s})$ and 
$\sigma^2_{\epsilon_s}$ from the hospitalization model in 
(\ref{eq:final_hosp}) were considered the same across all seasons so that 
all $\boldsymbol{\alpha}_s = \boldsymbol{\alpha}$. Note that this is not
a model assumption we make but only a simplyfying assumption for simulating
the data.
The values for $\boldsymbol{\alpha}$, $\sigma^2_{\epsilon}$, and $\phi$ were 
estimated by fitting model (\ref{eq:final_hosp}) using ILI and hospitalization 
data from the 2022 season. For fitting, the hospitalization data was first 
log-transformed. We took $\boldsymbol{\alpha}_{\phi} = (\boldsymbol{\alpha}, \phi)$
and assigned the noninformative prior 
$p(\boldsymbol{\alpha}_{\phi}, \sigma^2_{\epsilon}) \propto 1/\sigma^2_{\epsilon}$. 
The marginal posterior distribution 
$\boldsymbol{\alpha}_{\phi} | \sigma^2_{\epsilon}, \boldsymbol{H_{22}}$ 
was then the established posterior multivariate normal distribution and 
$\sigma^2_{\epsilon} | \boldsymbol{H_{22}}$ the inverse-$\chi^2$ posterior 
distribution \cite[]{gelman2013bayesian}. The posterior means of those 
parameters were used as the values from which log-hospitalizations were 
simulated. The number of replicates in the simulation was 500.

Model comparison was done by calculating a proper scoring rule for each 
forecast.
Proper scoring rules are the 
current standard for comparing performance between probabilistic forecasts 
and selecting the best forecasts according to the notion of maximizing 
sharpness subject to (auto-)calibration 
\cite[]{gneiting2007probabilistic, tsyplakov2013evaluation}. Proper scoring 
rules are commonly used in forecast comparison, and they are designed such 
that a 
forecaster is incentivized to be honest in the reporting of their forecasts 
\cite[]{gneiting2007strictly, gneiting2014probabilistic}.
The proper scoring rule primarily used in the FluSight competition and the 
COVID-19 Forecast Hub, and which is used in this simulation study,
is the negatively oriented
(smaller is better)
weighted interval score (WIS)
\cite[]{mathis2024evaluation, bracher2021evaluating}.
The WIS is used for scoring quantile or interval 
forecasts, or forecasts made up of predicitive intervals of several nominal 
levels
\cite[]{gneiting2007strictly, gneiting2014probabilistic, bracher2021evaluating}. 
The WIS defined in (\ref{eq:wis2}) where $Q$ is a forecast represented by all 
included quantiles, $B$ is the number of intervals,  $y^*$ is the observed 
value targeted by the forecast, $w_0 = 1/2$ and $w_b = \alpha_b / 2$ are 
weights for each interval, and $\alpha_b$ is the nominal level of the $b^{th}$ 
interval where $b \in \{1, ..., B\}$. 
$IS_{\alpha}$ is the interval score (IS), a proper scoring rule for a 
single interval, defined in (\ref{eq:is}). Here 
$\vmathbb{1}\{\cdot\}$ is the indicator function.

\begin{equation}
\label{eq:wis2}
        WIS_{0,B}(Q, y^*) = \frac{1}{B + 1/2} \times (w_0\times |y^* - median| + \sum_{b=1}^B \{w_k \times IS_{\alpha_b}(Q, y^*) \} )
\end{equation}

\begin{equation}
\label{eq:is}
        IS_{\alpha}(l,r;y^*) = (r-l) + \frac{2}{\alpha}(l - y^*)\vmathbb{1}\{y^* < l\} + \frac{2}{\alpha}(y^* - r) \vmathbb{1}\{y^* > r\}
\end{equation}

%

\begin{figure}[hbt!]
    \centering
    \includegraphics[scale=.45]{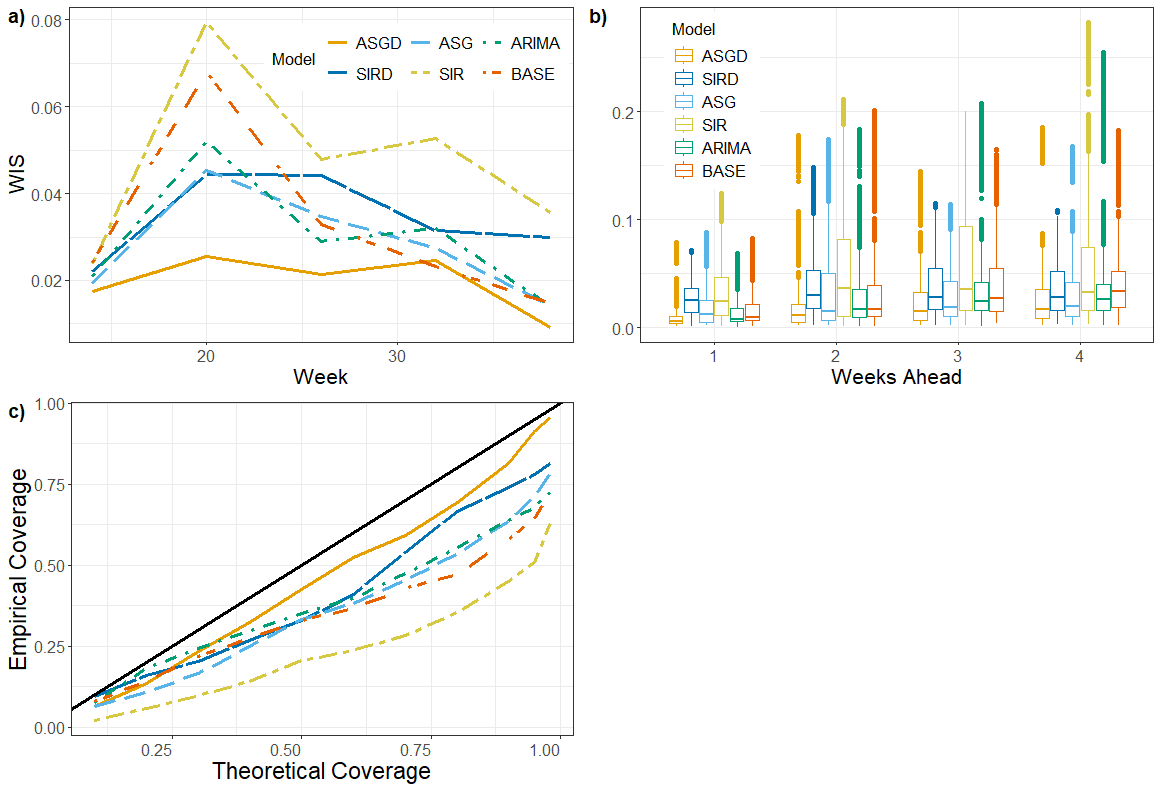}
    \caption{Results of forecasts for simulated hospitalizations for ASGD, ASG,
    SIRD, SIR, BASE, and ARIMA models. Plot a), WIS averaged over all
    left out seasons and 500 simulation replicated for each of weeks 14, 20, 
    26, 32, and 38, coloured by model. Plot b), boxplots of all WIS 
    scores separated by forecast horizon of 1 to 4 weeks ahead. Plot c),
    empirical calibration of each forecast model as the overall 
    forecast coverage of predictive intervals for the 11 nominal levels
    10\%, 20\%, ..., 90\%, 95\%, 98\%.}
    \label{fig:wis_by_week_horizon}
\end{figure}

The forecasts used in this section and in the subsequent section follow the 
FluSight convention of being quantiles corresponding to the 23 given 
probabilities $\\(0.010, 0.025, 0.050, 0.100, 0.150, …, 0.950, 0.975, 0.990)$
making each forecast a set of 11 predictive intervals and a predictive median.
Figure \ref{fig:wis_by_week_horizon} gives three plots showing results of the 
study. Plot a) shows WIS scores averaged over each left out season and each 
of the 500 simulation replicates. The ASGD forecasts on average clearly 
outperformed the other models considered in almost every week. The ASGD does
especially well at week 20 which contains forecasts of the holiday week 22. 
This demonstrates the
particular importance of modeling discrepancy at and around that 
week. The SIRD outperforms the SIR model, again showing the importance of 
modeling discrepancy, but it is still outperformed by the ASG and even for a
few weeks by the BASE and ARIMA models. Plot b) shows boxplots of all forecast
scores by model and separated by horizon forecasts of 1 to 4 weeks ahead. Here
also the ASGD performs the best with the other variations of the 
nonlinear models not always outperforming the BASE or ARIMA models. 
To assess calibration, plot c)
shows the empirical coverage of each model for 11 predictive intervals from 
10\% to 98\% nominal levels. Each model is undercalibrated, but the ASGD 
model is the closest to being well calibrated, especially at the hightest 
nominal levels. Table \ref{tab:sim_wis_res} shows summaries of the mean WIS and
coverage for 50\% and 95\% predictive intervals.


\begin{table}[ht]
\centering
\caption{Summary values of forecast performance in simulation study for 
    ASGD, ASG,
    SIRD, SIR, BASE, and ARIMA models. WIS, 50\% and 95\% predictive interval
    coverage averaged over all seasons and 500 simulation replicates.}
\begin{tabular}{lrrr}
Model & WIS & 50\%Coverage  & 95\% Coverage \\ 
  \hline
ASGD & 0.02 & 0.42 & 0.91 \\ 
  ASG & 0.03 & 0.33 & 0.71 \\ 
  ARIMA & 0.03 & 0.35 & 0.68 \\ 
  BASE & 0.03 & 0.33 & 0.65 \\ 
  SIRD & 0.03 & 0.33 & 0.78 \\ 
  SIR & 0.05 & 0.20 & 0.51 \\
\end{tabular}
\label{tab:sim_wis_res}
\end{table}

The results of this simulation study demonstrate the effectiveness of using the
assymetric Gaussian function along with modeling discrepancy for modeling
ILI and making flu hospitalization
forecasts in the US. The following section shows similar effectiveness of
coupling an effective ILI flu model with a hospitalization model to forecast
hospitalizations during the 2023 flu season.

\section{Analysis of forecasts for 2023 flu season}
\label{sec:analysis}

In this section we apply the forecast models to make forecasts for the 2023 
flu season weekly hospitalizations and compare the results with those of 
forecasts submitted to FluSight. The scoring of the forecasts is in the 
context of the FluSight competition where each competing forecast was 
submitted as a set of 23 quantiles corresponding to given probabilities.
Forecasts of 1, 2, 3, and 4-week ahead hospitalization counts were requested, 
and forecasts were made at the state and national levels. The first week of 
forecasting took place during the week of October 7, 2023, and the final week 
was the week of May 4, 2024 making 30 total weeks of forecasts.
 The same format was used during the 2021 and 2022 seasons and for the 
 COVID-19 Forecast Hub \cite[]{mathis2024evaluation, bracher2021evaluating}. 
 Primary scores for evaluating each forecast were the WIS and the 
 log-weighted intveral score (LWIS). The LWIS is defined the same as the WIS
 but where the log of the observation as well as the 
 log forecast quantiles are taken
 prior to calculating the score. We primarily use the LWIS in this section,
 though we including results for the WIS in the 
 supplementary materials.

Figure \ref{fig:normal_flu_forecasts} shows hospitalization forecasts at 
four different weeks during the season for ASGD, ASG, SIRD, and SIR ILI models
where hospitalizations are modeled according to model (\ref{eq:final_hosp})
where ILI is a quadratic predictor of hospitalizations, and hospitalizations
are assumed to be normally distributed. Of these forecasts, the ASGD appears 
to do the best job at capturing the trajectory of hospitalizations as well
as most frequently
making the sharpest predictions, particularly later in the season. 

\begin{figure}[hbt!]

  \centering
  \includegraphics[scale=.45]{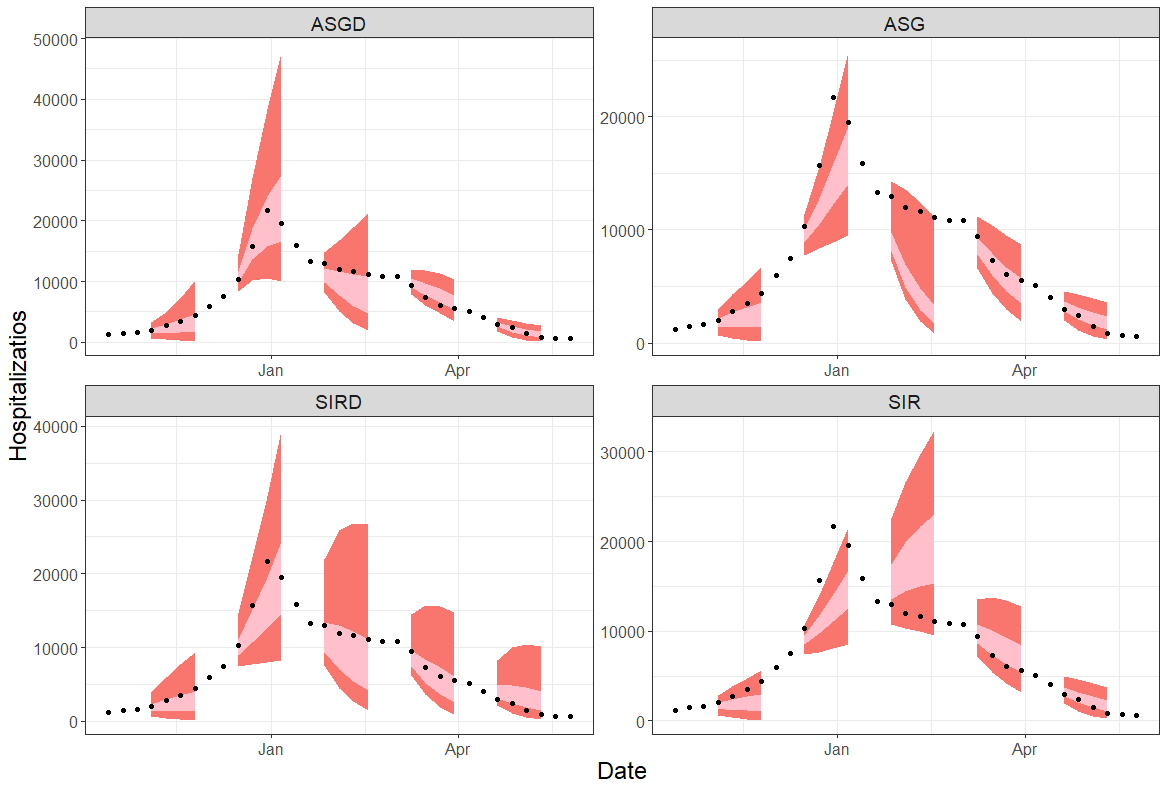}
\caption{Forecasts 1-4 weeks ahead for US hospitalizations during the 2023 
season for weeks 14, 20, 26, and 32. Forecasts are separated by ILI model, 
and the hospitalization models are all normally distributed and ILI is a 
quadratic predictor.}
\label{fig:normal_flu_forecasts}
\end{figure}

For the
remainder of this section, focus will be made on forecasts made by coupling the
ASGD and SIRD models with the 
normal hospitalization model where ILI is a quadratic predictor.
These forecast schemes are denoted as ASGD\_NORM2 and SIRD\_NOMR2. 
In the supplementary materials, we include analysis of 2023 hospitalization
forecasts for several other forecast
models including ASG and SIR ILI models coupled with other variations of the
hospitalization model where the ASGD\_NORM2 showed the best results
and the SIRD\_NORM2 performed nearly as well as the ASGD\_NORM2.

The FluSight 
models which are used for comparison are the
20 non-ensemble models
submitted to FluSight during the 2023 season.
The submitted forecasts may be found at \cite{mathis2023flusight}.
Among the models
was the FluSight baseline. The baseline forecast had as a median the most 
recently observed 
hospitalization count, and the forecast uncertainty was based on 
differences between 
previous hospitalizations. The baseline forecast model was
similar to the baseline forecast model used in
previous flu forecasting seasons and in the COVID-19 hub 
\cite[]{mathis2024evaluation, Cramer2022-hub-dataset}.

\begin{figure}[hbt!]
    
    \centering
    \includegraphics[scale = .6]{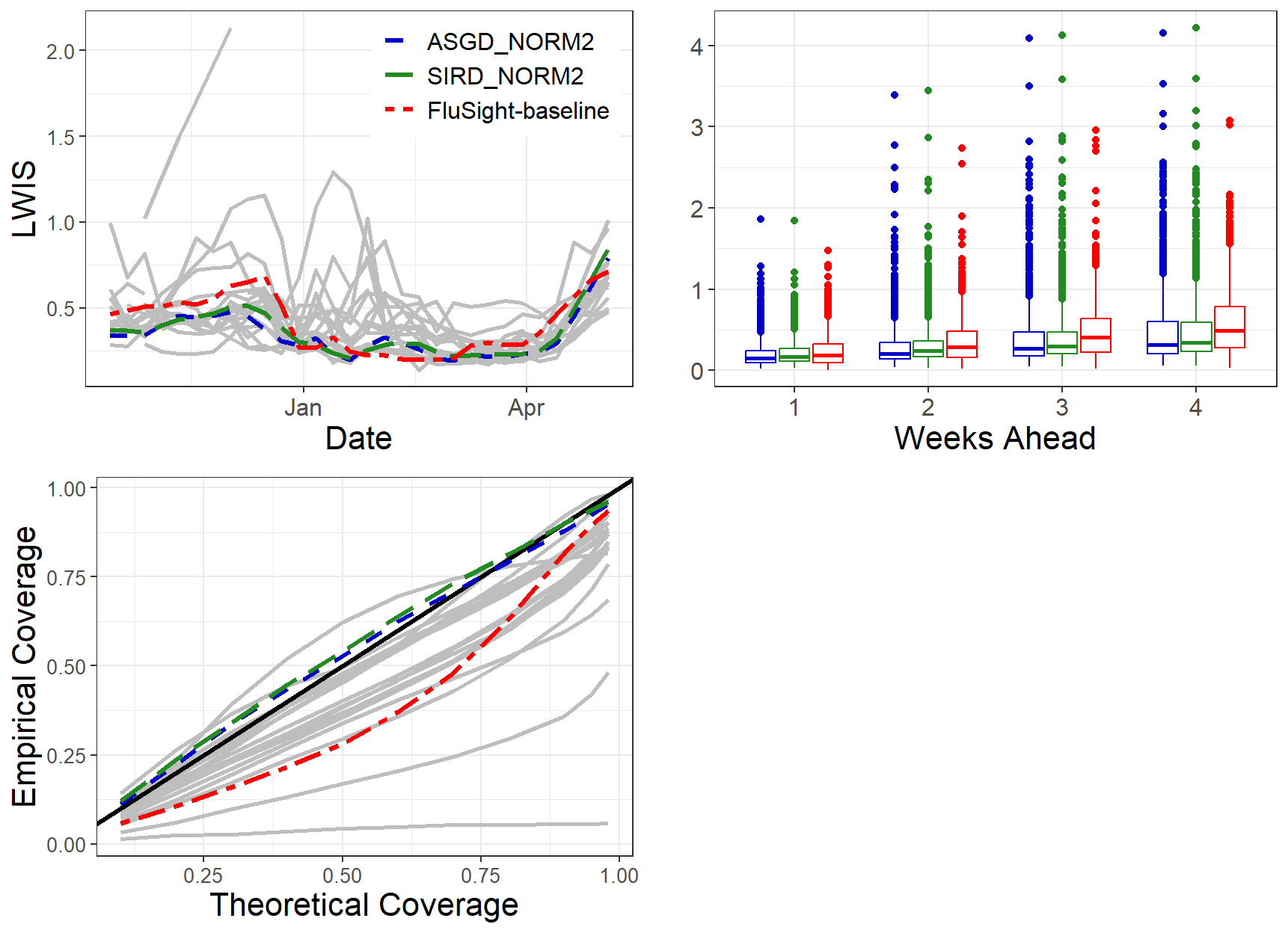}
    \caption{a) Log-weighted interval score (LWIS) averaged 
    over location and horizon for each week during the season for the 21 models being compared. 
    Each grey line represents one model, but the ASGD\_NORM2 (blue), 
    SIRD\_NORM2 (green), and 
    FluSight-baseline (red) are colored to stand out.
    b) Boxplots of all LWIS scores for 1, 2, 3, and 4-week ahead horizons 
    for the ASGD\_NORM2 (blue) and FluSight-baseline (red) models.
    c) Plot of empirical coverage against theoretical coverage for 11 
    predictive intervals from the 10\% nominal level to the 98\%. The black
    line $x = y$ shows where optimal coverage would fall.
    Each grey line represents one model, but the ASGD\_NORM2 (blue),
    SIRD\_NORM2 and 
    FluSight-baseline (red) are colored to stand out.}
    \label{fig:lwis_cover_sum}
\end{figure}

Figure \ref{fig:lwis_cover_sum} shows forecast results for the 2023 flu 
hospitalization forecasts for all 22 models being compared but with results
for the ASGD\_NORM2, SIRD\_NORM2, and FluSight baseline models highlighted. 
For each week of the 2023 season, the  
LWIS averaged over location and horizon 
for each model is shown in a). Boxplots of LWIS for 1, 2, 3, and 4-week ahead
horizons for the ASGD\_NORM2, SIRD\_NORM2, and baseline models are shown in b). 
And 
overall empirical coverage of the 11 predictive intervals for all models is 
shown in c) with ASGD\_NORM2, SIRD\_NORM2
and the baseline model highlighted. 
These plots show that over the season the ASGD\_NORM2 model tended to 
outperform most others in terms of LWIS, and in general is superior to the 
baseline model. The coverage plot shows that the ASGD\_NORM2 model is well 
calibrated at most levels, with slight overcoverage at the lower levels
up to about 80\%, and has better coverage than all other models. Similar results
for the WIS are shown in the supplementary materials.

\begin{figure}[hbt!]
    \centering
    \makebox[\linewidth][l]{
    \includegraphics[scale = .5]{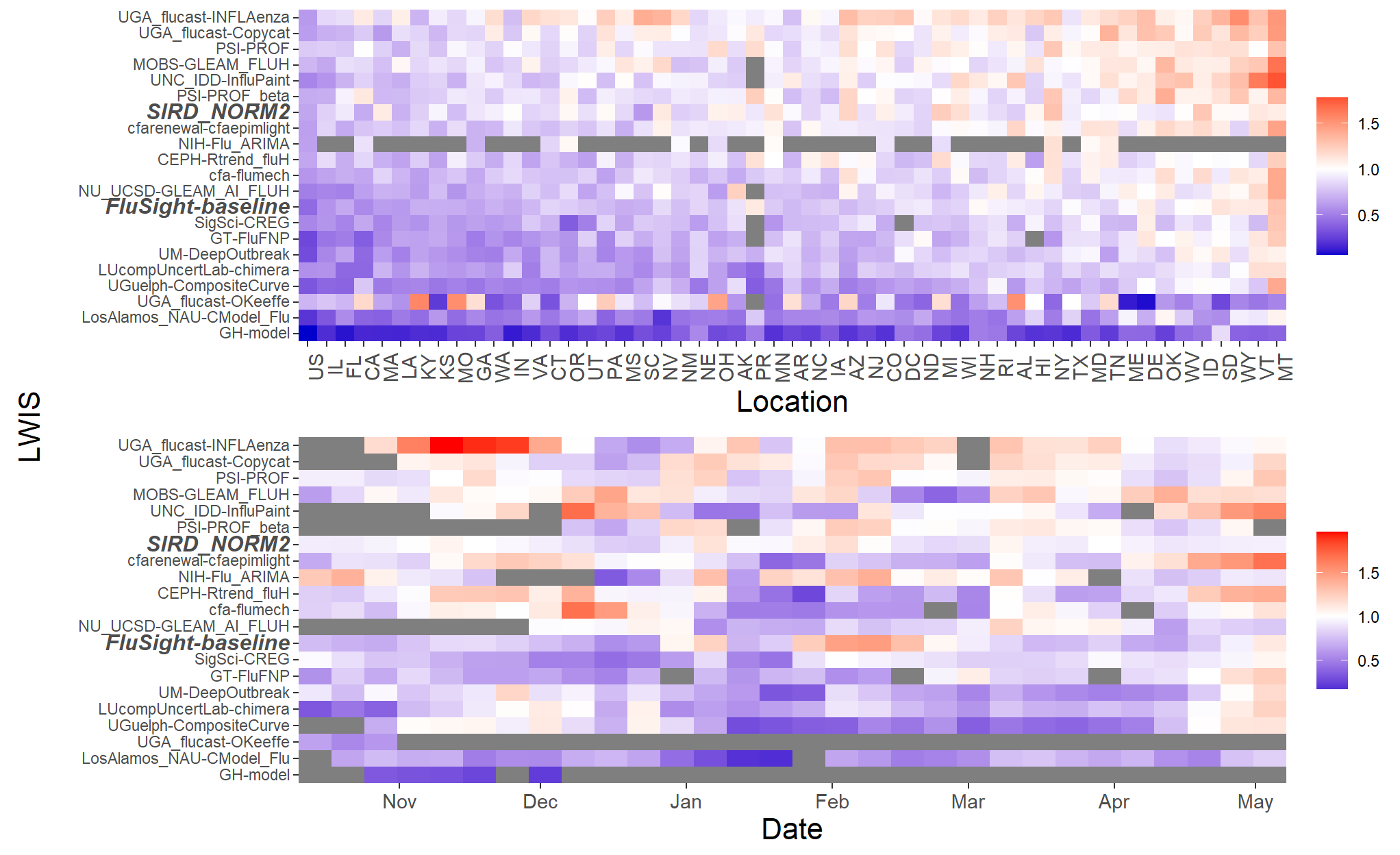}
    }
    \caption{Ratio of average log-weighted interval score (LWIS) for ASGD\_NORM2 
    to that of each competing FluSight model. Scores are averaged
    over all targets within indicated location (top) or week (bottom) and 
    only include forecasts shared by both models. Lower scores (blue) indicate
    superior performance by the ASGD\_NORM2 model and higher scores (red)
    indicate superior performance by competing model.}
    \label{fig:state_and_date_lwis}
    
\end{figure}

During the FluSight competition, most participating teams would occasionally 
miss a forecast submission for some week, location, or horizon. This makes a 
simple average of scores an imperfect measure for comparison, as forecasts
of certain targets included for one model may be excluded by another model.
Thus in figure \ref{fig:state_and_date_lwis} 
the ASGD\_NORM2 model is compared with all 
other models individually using LWIS scores averaged over only the 
forecasts shared by the two models.
Each tile represents the ratio of average ASGD\_NORM2
LWIS to that of one FluSight model by location or by week.
We call this the relative LWIS (RLWIS). A value smaller than
1 indicates the ASGD\_NORM2 model outperforms the competing model. A value 
higher than 1 indicateds the ASGD\_NORM2 is outperformed. A grey tile indicates
the competing model had no forecasts matching the specified target.
Similar plots showing WIS scores are found in the supplementary materials. 
The ASGD\_NORM2 outperforms most other competing models more often than not, but
there are some weeks or locations where forecasting was poorer. For example, 
the plot shows the ASGD\_NORM2 performed poorly in Montana relative to most other
models.

Table \ref{tab:fin_analysis_stats} gives summary results for mean and median
WIS and LWIS scores, predictive coverage, and relative RWIS and RLWIS scores
for each model. Along with the plots in this section, it 
shows that over the course of the 
2023 season, 
the ASGD\_NORM2 outperformed every other model except the 
UGA\_flucast-INFLAenza. The SIRD\_NORM2 is also one of the top performing 
forecast models.

\begin{table}[ht]
\centering
\caption{Overall summary scores for each of the 20 non-ensemble FluSight
models and the ASGD\_NORM2 model. Summaries include mean weighted interval
and log-weighted interval scores (MWIS/MLWIS), 
median weighted interval and log-weighted interval scores (MedWIS/MedLWIS), 
mean squared error
difference between predictive model empirical and theoretical coverage (MSEC)
see figure \ref{fig:lwis_cover_sum} c)), percentage of forecasts for all 
targets during the 2023 season (\% Forcs), and relative weighted interval
and log-weighted interval scores (RWIS/RLWIS).}
\small
\begin{tabular}{lrrrrrrrr}
 & MWIS & MLWIS & MedWIS & MedLWIS & MSEC & \% Forcs & RWIS & RLWIS\\ 
  \hline
ASGD\_NORM2 & 62 & 0.34 & 14 & 0.23 & 0.001 & 100 & 1 & 1 \\ 
  PSI-PROF & 63 & 0.35 & 15 & 0.23 & 0.003 & 100 & 0.90 & 0.97 \\ 
  SIRD\_NORM2 & 67 & 0.36 & 16 & 0.26 & 0.001 & 100 & 0.93 & 0.95 \\
  CEPH-Rtrend\_fluH & 69 & 0.39 & 13 & 0.28 & 0.004 & 100 & 0.81 & 0.86 \\ 
  FluSight-baseline & 85 & 0.41 & 16 & 0.32 & 0.023 & 100 & 0.63 & 0.82 \\ 
  UM-DeepOutbreak & 83 & 0.46 & 21 & 0.39 & 0.010 & 99 & 0.64 & 0.72 \\ 
  cfarenewal-cfaepimlight & 76 & 0.37 & 15 & 0.26 & 0.001 & 99 & 0.72 & 0.92 \\ 
  MOBS-GLEAM\_FLUH & 66 & 0.35 & 15 & 0.22 & 0.021 & 98 & 0.87 & 0.95 \\ 
  LosAlamos\_NAU-CModel\_Flu & 242 & 0.64 & 24 & 0.42 & 0.153 & 93 & 0.21 & 0.53 \\ 
  UGuelph-CompositeCurve & 132 & 0.52 & 25 & 0.41 & 0.018 & 93 & 0.43 & 0.62 \\ 
  LUcompUncertLab-chimera & 84 & 0.43 & 17 & 0.33 & 0.004 & 92 & 0.67 & 0.71 \\ 
  UGA\_flucast-INFLAenza & 74 & 0.32 & 15 & 0.20 & 0.003 & 90 & 0.77 & 1.07 \\ 
  cfa-flumech & 101 & 0.41 & 17 & 0.27 & 0.019 & 89 & 0.58 & 0.85 \\ 
  UGA\_flucast-Copycat & 76 & 0.35 & 16 & 0.21 & 0.012 & 87 & 0.81 & 0.98 \\ 
  GT-FluFNP & 105 & 0.48 & 18 & 0.38 & 0.046 & 86 & 0.45 & 0.73 \\ 
  UNC\_IDD-InfluPaint & 107 & 0.35 & 16 & 0.24 & 0.043 & 78 & 0.58 & 0.95 \\ 
  NU\_UCSD-GLEAM\_AI\_FLUH & 98 & 0.39 & 22 & 0.24 & 0.014 & 74 & 0.68 & 0.83 \\ 
  SigSci-CREG & 59 & 0.49 & 12 & 0.31 & 0.018 & 71 & 0.62 & 0.74 \\ 
  PSI-PROF\_beta & 77 & 0.32 & 18 & 0.20 & 0.004 & 67 & 0.87 & 0.93 \\ 
  NIH-Flu\_ARIMA & 230 & 0.34 & 28 & 0.23 & 0.001 & 20 & 0.57 & 0.88 \\ 
  GH-model & 161 & 1.52 & 28 & 1.47 & 0.373 & 17 & 0.18 & 0.27 \\ 
  UGA\_flucast-OKeeffe & 27 & 0.59 & 8 & 0.43 & 0.003 & 10 & 0.60 & 0.58 \\  
\end{tabular}
\label{tab:fin_analysis_stats}
\end{table}

\section{Conclusion}
\label{sec:conclusion}

 In this manuscript we introduce a statistical modeling framework which allows 
 for the incorporation of several nonlinear
 ILI forecast modeling methods. Specifically, 
 we built upon Osthus et al. \cite[]{osthus2019dynamic} and introduced a 
 framework for modeling ILI which includes the use of an arbitrary function 
 for modeling the main trajectory of ILI along with modeling the discrepancy and
 specifically incorporate the smooth ASG function.
 We model flu hospitalizations by incorporating the ILI forecast model into a 
 model forecasting hospitalizations where hospitalization predictions are a 
 linear or quadratic function of ILI. Utlizing the ILI data which has been
 available for many more years than flu hospitalization, greatly improves
 one's ability to forecast hospitalizations, and the borrowing information 
 from all previous seasons through hierarhcical modeling is able to further
 exploit the ILI data. Including model discrepancy was an important piece of 
 successful forecasting, but it required making thoughtful model constraints
 and selection of prior distributions.

The simulation study in section \ref{sec:simulation2} suggests the ASGD 
function in ILI modeling outperforms the SIRD model and out of the box 
forecast methods according to 
the WIS scoring rule, and in the analysis of 2023 FluSight forecasts the 
ASGD\_NORM2 modeling scheme outperformed all but one of the competing
forecasts when looking at the whole season. Though the ASGD\_NORM2 did very 
well across the whole season, there were a small number of weeks or locations
where performance was not as good.
It should not be assumed that ASGD\_NORM2 forecasts would perform
as well in every season given that each flu season will come with its own
challenges which the modeling scheme may or may not address.

The models in this manuscript were such that the data in each
location were modeled as
independent from all other locations. It is reasonable to assume that a
disease outbreak in one state will share characterstics with the outbreak in 
neighboring states, and indeed the behavior of ILI and hospitalizations is more
closely related for neighboring states as shown in data plots in the
supplementary materials. Notably, the one 2023 FluSight forecast model which 
outperformed
the ASGD\_NORM2, the UGA\_flucast-INFLAenza, was listed as being a spatial
model. And in past seasons, some of the top performing forecast models also 
explicitly modeled spatial relationships \cite[]{osthus2021multiscale}.
Likewise, the forecasts to which ours was compared were all non-ensemble 
forecasts which were all outperformed by the ensemble forecasts in FluSight.
Futher extensions of the models herein could include explicit modeling of 
spatial relationships between states and inclusion in an ensemble of other
forecast models.


\bibliographystyle{plainnat}
\bibliography{./master_bib}

\end{document}